\RequirePackage[2020-02-02]{latexrelease}
\documentclass[12 pt,twoside,aps,prd,amsmath,amssymb,
tightenlines,showpacs,showkeys,eqsecnum]{revtex4}
\UseRawInputEncoding
\usepackage{graphicx}
\usepackage{mathptmx}
\usepackage{bm}
\usepackage{amssymb}
\usepackage{amsmath}

\newcommand {\cG}{\cal G}
\newcommand {\cD}{\cal D}
\newcommand {\bg}{\bar \gamma}
\newcommand {\bp}{\bar \psi}

\def\myfigure #1#2#3#4
{\begin{figure}[ht]\begin{center}
\includegraphics[width=#2 \textwidth]{#1.jpg}
\parbox[t]{#4\textwidth}{\caption{#3}\label{#1}}
\end{center}\end{figure}}

\def \myfigures #1#2#3#4#5#6#7#8
{\begin{figure}[ht]
    \begin{center}
        \includegraphics[width=#2 \textwidth]{#1.jpg}
        \hfill
        \includegraphics[width=#6 \textwidth]{#5.jpg}
        \parbox[t]{#4\textwidth}{\caption {#3}\label{#1}}
        \hfill
        \parbox[t]{#8\textwidth}{\caption {#7}\label{#5}}
    \end{center}
\end{figure} }

\begin{document}

\baselineskip -24pt
\title{Maxwell-Dirac system in cosmology}
\author{Bijan Saha}
\affiliation{Laboratory of Information Technologies\\
Joint Institute for Nuclear Research\\
141980 Dubna, Moscow region, Russia\\ and\\
Peoples' Friendship University of Russia (RUDN University)\\
6 Miklukho-Maklaya Street, Moscow, Russian Federation\\
orcid: 0000-0003-2812-8930 } \email{bijan@jinr.ru}
\homepage{http://spinor.bijansaha.ru}

\hskip 1 cm

\begin{abstract}
Within the scope of a Bianchi type-I (BI) cosmological model we
study the interacting system of spinor and electromagnetic fields
and its role in the evolution of the Universe. In some earlier
studies it was found that in case of a pure spinor field the
presence of nontrivial non-diagonal components of EMT leads to some
severe restrictions both on the spacetime geometry and/or spinor field
itself, whereas in case of electromagnetic field with induced
nonlinearity such components impose severe restrictions on metric functions
and the components of the vector potential. It is shown that in case of interacting
spinor and electromagnetic fields restrictions are not as severe as
in other cases and in this case a nonlinear and massive spinor field with different
components of vector potential can survive in a general Bianchi type-I spacetime.
\end{abstract}



\maketitle

\bigskip

\section{Introduction}

The present day Universe wonderfully homogeneous and isotropic. This
fact convinces most cosmologists perform their studies in the framework of
Friedmann-Lamaitre-Robertson-Walker (FLRW) cosmological model. But at the same time
there are both theoretical and observational arguments for an
anisotropic phase of the universe which becomes isotropic in the course of evolution \cite{misner,WMAP,Hinshaw}.
Since Bianchi type-I (BI) is the simplest generalization of the FLRW spacetime, many
authors considered this model, which in their view can shed some
light on the early anisotropy of the Universe. In a number of papers
we have studied role of nonlinear spinor field in the evolution of a
initially anisotropic Universe
 \cite{Saha1997GRG,Saha1997JMP,SahaPRD2001,greene,Saha2004bPRD,Saha2006PRD,FabGRG,ELKO,PopPRD}.
Further study suggests that the introduction of spinor field into the system gives rise to the nontrivial non-diagonal components of the energy-momentum tensor (EMT) \cite{2018EChAYa49,2015APSS357,uniserse2023}. These non-diagonal components impose different
kinds of restrictions both on spacetime geometry and spinor fields
itself. In case of a BI cosmological model we are left with three
options: (i) the spinor fields becomes linear and massless; (ii)
spacetime becomes isotropic or (iii) BI spacetime transfers to a
locally rotationally symmetric (LRS-BI) one with some restrictions
on the spinor field. On the other hand if the BI spacetime is filled
with electromagnetic field given by $A_\mu = \{0,\,A_1 (t),\,
A_2(t),\,A_3(t)\}$, the presence of non-diagonal EMT leads to $A_1 =
A_2 = A_3$ and the spacetime becomes isotropic \cite{2011CEJP9-1165-1172}.

It should be noted that the Maxwell-Dirac system is well studied in mathematical physics. The corresponding Maxwell-Dirac system is given by \cite{georgiev}
\begin{subequations}
\begin{align}
    4 \pi \partial_\mu \partial^\mu A^\nu &= j^\nu, \label{Max}\\
    \left(\imath\bg^\mu \partial_\mu - \bg^\mu A_\mu \right) \psi - m \psi &= 0, \label{dir1}
\end{align}
\end{subequations}
with the Lorentz condition
\begin{align}
    \partial_\mu A^\mu = 0, \label{Lorentz}
\end{align}
with $j^\nu = \bp \gamma^\nu \psi$. Such system can be obtained from the Lagrangian with the interacting term
\begin{align}
    L_{int} = A_\mu J^\mu. \label{int0}
\end{align}
Exact solutions of Maxwell-Dirac equations were investigated in
\cite{Das1,Das2}.

\section{Basic Equations}

Let us consider a system of interacting spinor and electromagnetic fields within the scope of an anisotropic but homogeneous Bianchi type-I (BI) gravitational cosmological model. The corresponding action we choose in the form
\begin{align}
    {\cal S}(g; \psi, \bp) = \int\, \left(L_{\rm g} +
        L_{\rm sp}+ L_{\rm em} + L_{\rm int} \right) \sqrt{-g} d^4 x.    \label{action}
\end{align}
Here $L_{\rm g}$ corresponds to the gravitational field,
\begin{align}
        L_{\rm g} = \frac{R}{2\kappa},       \label{lgrav}
\end{align}
where  $R$ is the scalar curvature,  $\kappa = 8 \pi G$ with $G$ being Newton's
  gravitational constant.

The spinor field Lagrangian $L_{\rm sp}$ is given in the form \cite{SahaPRD2001}
\begin{align}
    L_{\rm sp} = \frac{\imath}{2} \left[\bp \gamma^{\mu} \nabla_{\mu}
            \psi- \nabla_{\mu} \bar \psi \gamma^{\mu} \psi \right] - m \bp \psi - \lambda_1 Y(K)
             \label{lspin}
\end{align}
Here $m$ is the spinor field mass, $Y$ is a self-coupling function that can be
  positive or negative, and $K = b_1 I + b_2 J$, where  $b_1$ and $b_2$
  take the values $1, 0, -1$, so that $K$ may be one of the following expressions:
\begin{align}                               \label{IJK}
     K = \{I,\,J,\,I+J,\,I-J\},\qquad
     I = S^2 = (\bp \psi)^2, \qquad   J = P^2 = (\imath \bp \bg^5 \psi)^2.
\end{align}
According to Fierz identity $Y(K)$ describes the spinor field nonlinearity in its most general form \cite{fierz}.

The electromagnetic field Lagrangian is taken in the conventional form
\begin{align}
        L_{\rm em} = -\frac{1}{16 \pi} F_{\tau \eta} F^{\tau \eta},         \label{em}
\end{align}
  while the interaction Lagrangian is chosen as
\begin{align}
    L_{\rm int} =  -\frac{\lambda_2}{16 \pi} F_{\tau \eta} F^{\tau \eta} Z(K),    \label{int}
\end{align}
Like in \eqref{lspin} $Y$, in \eqref{int} $Z$ is some arbitrary function of $K$.
  For convenience we further combine \eqref{em} and \eqref{int} to write
\begin{align}
        L_{\rm emint} = -\frac{1}{16 \pi} F_{\tau \eta} F^{\tau\eta} X (K), \quad
        X(K) \equiv 1 + \lambda_2 Z(K).  \label{emint}
\end{align}
The spinor field equations corresponding to the Lagrangian \eqref{lspin} and \eqref{int} are
\begin{subequations}                 \label{speq}
\begin{align}
    \imath\gamma^\mu \nabla_\mu \psi - m \psi - {\cD} \psi
            - \imath {\cG} \gamma^5 \psi &= 0,                  \label{speq1}
\\
    \imath \nabla_\mu \bp \gamma^\mu +  m \bp + {\cD}\bp
            + \imath {\cG}  \bp \gamma^5 &= 0,                  \label{speq2}
\end{align}
\end{subequations}
where \begin{align}
        {\cD} = 2 b_1 S  \big[\lambda_1 Y_K + F_{\tau \eta} F^{\tau \eta} \lambda_2 Z_K /(16 \pi) \big],
\nonumber\\
        {\cG} = 2 b_2 P \big[\lambda_1 Y_K +  F_{\tau \eta} F^{\tau \eta} \lambda_2 Z_K /(16 \pi)\big],
        \nonumber
\end{align}
  with $F_K = dF/dK$ and $H_K = dH/dK$. In view of \eqref{speq} it can be shown that
\begin{align}
        L_{\rm sp} =  {\cal D} S + {\cal G} P - \lambda_1 Y =
        \lambda_1 \left(2 K Y_K - Y\right) +  F_{\tau \eta} F^{\tau \eta}\lambda_2 K Z_K/(8 \pi).      \label{LspinF}
\end{align}
  In the above expressions, $\nabla_\mu \psi = \partial_\mu \psi -\Omega_\mu \psi$
  and $\nabla_\mu \bp = \partial_\mu \bp + \bp \Omega_\mu$, where $\Omega_\mu$ is the spinor
  affine connection, defined by

  \begin{align}
\Omega_\mu = \frac{1}{4} \left(\bg_a \gamma^\beta \partial_\mu
e^{(a)}_\beta - \gamma_\rho \gamma^\beta \Gamma_{\mu
\beta}^\rho\right). \label{SPAC}
\end{align}
As one sees, the spinor affine connection is completely defined by the metric, hence spinor field becomes very sensitive to the gravitational one.

  The electromagnetic field equations take the form
\begin{align}
        \partial_\eta \left(\sqrt{-g} F^{\tau \eta} X (K)\right) = 0.                  \label{eqem}
\end{align}
  The total energy-momentum tensor (EMT) has the form
\begin{align}
    T_{\mu}^{\rho}&= \frac{\imath}{4} g^{\rho\nu} \Big(\bp \gamma_\mu \partial_\nu \psi
    + \bp \gamma_\nu \partial_\mu \psi - \partial_\mu \bar \psi \gamma_\nu \psi
     - \partial_\nu \bp \gamma_\mu \psi \Big) \nonumber\\
     &- \frac{\imath}{4} g^{\rho\nu} \bp \Big(\gamma_\mu \Omega_\nu +
        \Omega_\nu \gamma_\mu + \gamma_\nu \Omega_\mu + \Omega_\mu \gamma_\nu\Big)\psi
         - \delta_{\mu}^{\rho} \lambda_1 (2 K Y_K - Y(K)) \label{emt}  \\
   & - \frac{X(K)}{4\pi} \Big ( F_{\mu \eta} F^{\nu \eta}
            - \frac{1}{4} \delta_\mu^\nu F_{\tau\eta} F^{\tau \eta}\Big)
            - \frac{\lambda_2 K Z_K}{4\pi} \delta_\mu^\rho F_{\tau\eta} F^{\tau \eta}.                      \nonumber
\end{align}
The second term in \eqref{emt} plays crucial role as generally thanks to this term there occurs nontrivial non-diagonal components of EMT. Until now all the expressions for spinor and electromagnetic fields valid for any gravitational field. In what follows, we specify the gravitational field.

The gravitational field in our case is given by anisotropic cosmological BI model:
\begin{align}
ds^2 = dt^2 - a_1^2 {dx_1}^2 - a_2^2 {dx_2}^2 - a_3^2 {dx_3}^2,
\label{BI}
\end{align}
where the metric functions are the function of $t$ only, i.e., $a_i
= a_i(t)$. As one sees, it is the straight forward generalization of FLRW model, where $a_1 = a_2 = a_3 = a$

We will consider the case when electromagnetic 4-potential has only spatial components so that
$A_\mu = (0,\,A_1,\,A_2,\,A_3)$, and the spinor and the electromagnetic
fields depend on $t$ only, i.e., $\psi = \psi (t)$,\, $\bp = \bp
(t)$ and $A_i = A_i (t), \quad i = 1,\,2,\,3$. In this case
$F^{\mu\nu}$ has only the following non-zero components :
$F^{01},\,F^{02},\,F^{03}$.

In this case from \eqref{eqem} one dully finds
\begin{align}
F^{0i} = \frac{q_i}{V X(K)}, \quad F_{0i} = \dot A_i = g_{00} g_{ii} F^{0i} = -\frac{a_i^2q_i}{V X(K)} \quad q_i = {\rm const.}, \quad i = 1,\,2,\,3, \label{dotA}
\end{align}
where $V$ is the volume scale of the BI metric
\begin{align}
V = a_1 a_2 a_3. \label{Vsc}
\end{align}

To deal with spinor field we have to find the spinor affine connections corresponding to the metric \eqref{BI}. From \eqref{SPAC} in this case we find

\begin{align}
\Omega_0 = 0, \quad \Omega_1 =\frac{\dot a_1}{2}\bg^1 \bg^0, \quad
\Omega_2 = \frac{\dot a_2}{2}\bg^2 \bg^0, \quad \Omega_3 =\frac{\dot
a_3}{2}\bg^3 \bg^0. \label{CarSPAC}
\end{align}

The spinor field equations \eqref{speq} in this case read

\begin{subequations}                 \label{speqBI}
    \begin{align}
        \imath\gamma^0 \left(\dot \psi  + \frac{\dot V}{2 V} \psi\right) - m \psi - {\cD} \psi
        - \imath {\cG} \gamma^5 \psi &= 0,                  \label{speq1BI}
        \\
        \imath  \left(\dot \bp  + \frac{\dot V}{2 V} \bp\right) \gamma^0 +  m \bp + {\cD}\bp
        + \imath {\cG}  \bp \gamma^5 &= 0,                  \label{speq2BI}
    \end{align}
\end{subequations}

In view of \eqref{dotA} from \eqref{emt} in this case we find the
following nontrivial components of EMT of the interacting Dirac and
Maxwell system.

\begin{subequations}
\label{emtc}
\begin{align}
T_0^0 &= P_2 - Q
\left[q_1^2 a_1^2 + q_2^2 a_2^2 + q_3^2 a_3^2\right],
\label{emtc00}\\
T_1^1 &= P_1 + Q
\left[q_1^2 a_1^2 - q_2^2 a_2^2 - q_3^2 a_3^2\right],
\label{emtc11}\\
T_2^2 &= P_1 +Q
\left[- q_1^2 a_1^2 + q_2^2 a_2^2 - q_3^2 a_3^2\right],
\label{emtc22}\\
T_3^3 &= P_1 + Q
\left[-q_1^2 a_1^2 - q_2^2 a_2^2 + q_3^2 a_3^2\right],
\label{emtc33}\\
T_2^1 &= \frac{a_2 a_3}{4a_1} \left(\frac{\dot a_1}{a_1} -
\frac{\dot a_2}{a_2}\right) A^3 - \frac{q_1 q_2 a_2}{4\pi V^2 X},
\label{empc12}\\
T_1^3 &= \frac{a_1 a_2}{4a_3} \left(\frac{\dot a_3}{a_3} -
\frac{\dot a_1}{a_1}\right) A^2 - \frac{q_3 q_1 a_1}{4\pi V^2 X},
\label{empc13}\\
T_3^2 &= \frac{a_3 a_1}{4a_2} \left(\frac{\dot a_2}{a_2} -
\frac{\dot a_3}{a_3}\right) A^1 - \frac{ q_2 q_3 a_3}{4\pi V^2 X},
\label{empc23},
\end{align}
\end{subequations}
where we denote, $P_1 = \lambda_1 \left(Y - 2 K Y_K\right)$, $P_2 = m_{\rm sp} S + \lambda_1 Y$,
$Q = 1/(8\pi V^2 X)$ and $A^\mu = \bp \gamma^5 \gamma^\mu \psi.$ Note that $A$ with a
lower index ($A_\mu$) denotes a electromagnetic vector potential,
whereas, $A$ with an upper index ($A^\mu$) denotes a pseudovector
constructed from Dirac field. Let us now find the equations for the
invariants of the spinor field.

\begin{subequations}
\label{invs}
\begin{align}
\dot S_0^0 + 2 {\cG} A_0^0 &= 0, \label{S0}\\
\dot P_0^0 - 2 (m_{\rm sp} + {\cD}) A_0^0 & = 0, \label{P0}\\
\dot A_0^0 + 2 (m_{\rm sp} + {\cD}) P_0 - 2 {\cG} S_0 & = 0,
\label{A00}\\
\dot A_0^1 & = 0, \label{A10}\\
\dot A_0^2 & = 0, \label{A20}\\
\dot A_0^3 & = 0, \label{A30}
\end{align}
\end{subequations}
where we define $S_0 = S V$, $P_0 = P V$ and $A^\mu_0 = A^\mu V$. From \eqref{invs} one dully finds
\begin{align}
S^2 + P^2 + {A^0}^2  = \frac{c_0^2}{V^2}, \quad A^1 = \frac{c_1}{V},
\quad A^2  = \frac{c_2}{V}, \quad A^3
 = \frac{c_3}{V}, \label{inv}
\end{align}
where $c_0,\, c_1,\, c_2,\, c_3$ are constants. As it was shown
earlier, the invariant $K$ is related to the metric functions in the
following way \cite{SahaPRD2001}
\begin{align}
K = \frac{C^2}{V^2}, \quad C = const. \label{KV}
\end{align}
This relation is true for $K = \{J,\,I + J,\, I - J\}$ for a
massless spinor field, while for $ K = I$ it is valid both for
massive and massless spinor field.

Since the Einstein tensor corresponding to the metric \eqref{BI}
possesses only diagonal components, then  from \eqref{empc12},
\eqref{empc13} and \eqref{empc23} on account of \eqref{invs} one
finds
\begin{subequations}
\label{emtcn}
\begin{align}
\frac{\dot a_1}{a_1} - \frac{\dot a_2}{a_2} &= \frac{q_1 q_2
a_1}{\pi V X c_3 a_3}, \label{empc12n}\\
\frac{\dot a_2}{a_2} - \frac{\dot a_3}{a_3} &= \frac{q_2 q_3
a_2}{\pi V X c_1 a_1}, \label{empc23n}\\
\frac{\dot a_3}{a_3} - \frac{\dot a_1}{a_1} &=  \frac{q_3 q_1
a_3}{\pi V X c_2 a_2}. \label{empc13n}
\end{align}
\end{subequations}
Summation of \eqref{empc12n}, \eqref{empc13n} and \eqref{empc23n} on
account of \eqref{dotA} leads to
\begin{align}
c_1 c_2 q_1 q_2 a_1^2 a_2 + c_2 c_3 q_2 q_3 a_2^2 a_3 +
c_3 c_1 q_3 q_1 a_3^2 a_1 = 0. \label{rel1}
\end{align}
Note that the foregoing relation is valid both for interacting system spinor and electromagnetic fields, as well as for the system with minimal coupling.
It should be noted that in absence of electromagnetic field the non-diagonal components of EMT
\eqref{empc12}, \eqref{empc13} and \eqref{empc23} leads to \cite{2015APSS357}

\begin{subequations}
    \label{emtcn0}
    \begin{align}
        \left(\frac{\dot a_1}{a_1} - \frac{\dot a_2}{a_2}\right) A^3 &= 0, \label{empc12n1}\\
        \left(\frac{\dot a_3}{a_3} - \frac{\dot a_1}{a_1}\right) A^2 &= 0, \label{empc23n1}\\
        \left(\frac{\dot a_2}{a_2} - \frac{\dot a_3}{a_3}\right) A^1 &= 0, \label{empc13n1}
    \end{align}
\end{subequations}
which leads to three different cases:\\
(i) $A^1 = A^2 = A^3 = 0$ which leads to case with linear and massless spinor field;\\
(ii) $A^2 = A^3 = 0$ and $a_2 = a_3$ which leads to locally rotationally symmetric Bianchi type-I (LRSBI) model;\\
(iii) $a_1 = a_2 = a_3$ i.e. the BI spacetime becomes Friedmann-Lamaitre-Robertson-Walker (FLRW) spacetime. It should be emphasized that in an earlier paper we considered the electromagnetic field with induced nonlinearity in BI model that leads to the isotropization of the electromagnetic potential: $A_1 = A_2 = A_3$ \cite{2011CEJP9-1165-1172}.

The idea to introduce the electromagnetic field together with the spinor one was motivated by the fact whether it can allow a general BI spacetime with nonlinear spinor field. And as we see, it can.

Let us now solve the Diagonal equations of Einstein system

\begin{subequations}
\label{BIEbi}
\begin{align}
\frac{\ddot a_2}{a_2} +\frac{\ddot a_3}{a_3} + \frac{\dot
a_2}{a_2}\frac{\dot a_3}{a_3}&=  \kappa \left[ P_1  +  Q
\left[q_1^2 a_1^2 - q_2^2 a_2^2 - q_3^2 a_3^2\right]\right],\label{11bi}\\
\frac{\ddot a_3}{a_3} +\frac{\ddot a_1}{a_1} + \frac{\dot
a_3}{a_3}\frac{\dot a_1}{a_1}&= \kappa \left[P_1 + Q
\left[- q_1^2 a_1^2 + q_2^2 a_2^2 - q_3^2 a_3^2\right]\right],\label{22bi}\\
\frac{\ddot a_1}{a_1} +\frac{\ddot a_2}{a_2} + \frac{\dot
a_1}{a_1}\frac{\dot a_2}{a_2}&=  \kappa \left[P_1 + Q
\left[-q_1^2 a_1^2 - q_2^2 a_2^2 + q_3^2 a_3^2\right]\right],\label{33bi}\\
\frac{\dot a_1}{a_1}\frac{\dot a_2}{a_2} +\frac{\dot
a_2}{a_2}\frac{\dot a_3}{a_3} +\frac{\dot a_3}{a_3}\frac{\dot
a_1}{a_1}&= \kappa \left[P_2 - Q \left[q_1^2 a_1^2 + q_2^2 a_2^2 + q_3^2 a_3^2\right]\right].
\label{00bi}
\end{align}
\end{subequations}

Introducing directional Hubble parameter $H_i = \dot a_i/a_i$ we rewrite the foregoing system as follows:

\begin{subequations}
\label{BIEbiH2}
\begin{align}
\dot a_1 &= H_1 a_1, \label{H12}\\
\dot a_2 &= H_2 a_2, \label{H22}\\
\dot a_3 &= H_3 a_3, \label{H32}\\
\dot H_2 + \dot H_3 - H_2^2 - H_3^2 + H_2 H_3 &=  \kappa \left[ P_1
+ Q \left[q_1^2 a_1^2 - q_2^2 a_2^2 - q_3^2 a_3^2\right]\right],\label{11biH2}\\
\dot H_3 + \dot H_1 - H_3^2 - H_1^2 + H_3 H_1 &= \kappa \left[ P_1 +
Q \left[- q_1^2 a_1^2 + q_2^2 a_2^2 - q_3^2 a_3^2\right]\right],\label{22biH2}\\
\dot H_1 + \dot H_2 - H_1^2 - H_2^2 + H_1 H_2 &=  \kappa \left[ P_1
+ Q \left[-q_1^2 a_1^2 - q_2^2 a_2^2 + q_3^2 a_3^2\right]\right],\label{33biH2}\\
H_1 H_2 + H_2 H_3 + H_3 H_1 &= \kappa \left[P_2 - Q\left[q_1^2 a_1^2
+ q_2^2 a_2^2 + q_3^2 a_3^2\right]\right]. \label{00biH2}
\end{align}
\end{subequations}
It can be shown that \eqref{00biH2} is the consequence of three others, hence the foregoing system can be substituted by the following one:

\begin{subequations}
\label{BIEbiH3}
\begin{align}
\dot a_1 &= H_1 a_1, \label{H13}\\
\dot a_2 &= H_2 a_2, \label{H23}\\
\dot a_3 &= H_3 a_3, \label{H33}\\
\dot H_1 &= \frac{\kappa}{2} \left[P_1 + Q \left[q_3^2 a_3^2 + q_2^2
a_2^2 - 3 q_1^2 a_1^2\right]\right] + \frac{1}{2}\left[2 H_1^2 - H_1
H_2 - H_3 H_1 + H_2 H_3\right],\label{11biH3}\\
\dot H_2 &= \frac{\kappa}{2} \left[P_1 + Q \left[q_3^2 a_3^2 + q_1^2
a_1^2 - 3 q_2^2 a_2^2\right]\right] + \frac{1}{2}\left[2 H_2^2 - H_1
H_2 - H_2 H_3 + H_3 H_1\right],\label{22biH3}\\
\dot H_3 &= \frac{\kappa}{2} \left[P_1 + Q\left[q_2^2 a_2^2 + q_1^2
a_1^2  - 3 q_3^2 a_3^2 \right]\right] + \frac{1}{2}\left[2 H_3^2 -
H_3 H_1 - H_2 H_3 + H_1 H_2 \right].\label{33biH3}.
\end{align}
\end{subequations}
Since $S = C/V$ and $K = C^2/V^2$ and $V = a_1 a_2 a_3$, the right hand side of the equations \eqref{11biH3}, \eqref{22biH3} and \eqref{33biH3} are the functions of $H_i$ and $a_i$, only. In the
foregoing system $q_1, q_2, q_3$ are constants.

In order to solve the system \eqref{BIEbiH2} we have to give the
initial values for $a_i$'s and $H_i$'s, beside $q_i$'s. The relation
\eqref{rel1} is consistent with the system and can be exploited to
find say $c_3$ for the given values of other parameters at initial
time:

\begin{align}
 q_3 = -
\frac{c_1 c_2 q_1 q_2 a_1^2 a_2}{\left(c_2 c_3 q_2 a_2^2 a_3 +  c_3 c_1
 q_1 a_3^2 a_1\right)}. \label{q3}
\end{align}

In what follows we solve the system \eqref{BIEbiH3} numerically. For this we have to give $P_1$, $P_2$ and $Q$ explicitly. Let us assume that the functions $Y(K)$ and $Z(K)$ are power functions of $K$ such that $Y(K) = K^{n_1}$ and $Z(K)= K^{n_2}$. We consider a massive spinor field with $K = I$. Then on account of \eqref{Vsc} and \eqref{KV} we find  \\
$P_1 = \lambda_1 \left(Y - 2 K Y_K\right) = \lambda_1 \left(1 - 2 n_1\right) C^{2n_1}/(a_1 a_2 a_3)^{2n_1}$,\\
$P_2 = m_{\rm sp} S + \lambda_1 Y = m_{\rm sp} C/(a_1 a_2 a_3) + \lambda_1 C^{2n_1}/(a_1 a_2 a_3)^{2n_1}$, \\
and \\
$Q = 1/(8\pi V^2 X) = 1/(8 \pi (a_1 a_2 a_3)^2(1 + \lambda_2 C^{2 n_2}/(a_1 a_2 a_3)^{2 n_2}))$.

Our aim is to find some qualitative solution, so we set the values of parameters and initial conditions simple as possible. For example we set $c_1 = c_2 = c_2 = 1$ we also set $a_1(0) = a_2(0) = a_3(0) = 1$, $H_1(0) = H_2(0) = H_3(0) = 0.5$, $q_1 = 2$ and $q_2 = 3$. Then from \eqref{q3} we find $q_3 = -6/5$. The self-coupling term $\lambda_1 = 0.5$ and the interaction term $\lambda_2 = 0.7$. We also set $n_1 = 4$ and $n_2 = 3.$ Under these conditions we obtain the solutions for metric functions $a_i(t)$ and Hubble parameters $H_i(t)$ numerically.

\begin{figure}
    \centering
    \includegraphics[width=8 cm]{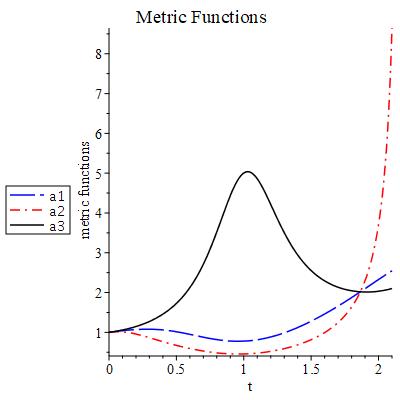}
    \caption{\label{Fig1}Evolution of the metric functions $a_1$, $a_2$ and $a_3$
    }
\end{figure}

\begin{figure}
    \centering
    \includegraphics[width=8 cm]{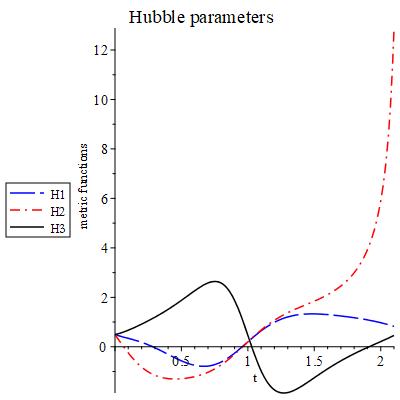}
    \caption{\label{Fig2}Evolution of the directional Hubble parameters $H_1$, $H_2$ and $H_3$}
\end{figure}

In figures \ref{Fig1} and \ref{Fig2} we plot the components of metric functions and Hubble parameter, respectively. In the figures blue long dash line stand for $x$-component, red dash dot line stands for $y$-component, and black solid line stand for $z$-component of metric function and Hubble parameter, respectively.

\section{conclusion}

Within the scope of an anisotropic BI cosmological model we study the role of an interacting system of spinor and electromagnetic fields in the evolution of Universe. Earlier, it was shown that if a BI spacetime is filled with nonlinear spinor field, the nontrivial non-diagonal components of the EMT impose some severe relations both on spacetime geometry and spinor field. In particular, the spinor field becomes massless and liner in BI geometry. In this report it is shown that the introduction of electromagnetic field allows massive and nonlinear spinor field in BI spacetime. The corresponding system is obtained and solved numerically.

 \vskip 1 cm

 \noindent {\bf Funding:} Not applicable

 \vskip .5 cm

 \noindent {\bf Institutional Review Board Statement:} Not applicable

 \vskip .5 cm

 \noindent {\bf Informed Consent Statement:} Not applicable

 \vskip .5 cm

 \noindent {\bf Acknowledgment:}\,\,\,\,{This paper has been
    supported by the RUDN University Strategic Academic Leadership
    Program.}

 \vskip .5 cm

 \noindent \textbf{DAS:} No datasets were generated or analyzed
 during the current study

 \vskip .5 cm

 \noindent {\bf Conflicts of Interest:} No conflict of interests

\end{document}